\documentstyle[epsf,epsfig,12pt]{article}
\mark{{}{}}
\def\nll{ \nonumber \\}

\def\lb{\left(}
\def\rb{\right)}

\def\z0{Z}
\def\gf{G_{\mu}}
\def\zm{M_{_Z}}

\def\hm{M_{_H}}
\def\wm{M_{_W}}

\def\barb{\overline b}

\def\barc{\overline c}

\def\barnu{\overline{\nu}}

\def\i3f{I^{(3)}_f}

\def\osp2{16\,\pi^2}
\def\ap2{\left(p^2\right)}

\def\tmo{\times 10^{-1}}
\def\tmt{\times 10^{-2}}

\def\s0h{\sigma^h_0}
\def\ba{\begin{eqnarray}}
\def\ea{\end{eqnarray}}

\def\beq{\begin{equation}}
\def\eeq{\end{equation}}
\def\bea{\begin{eqnarray}}
\def\eea{\end{eqnarray}}
\def\barr{\begin{array}}
\def\earr{\end{array}}
\def\bc{\begin{center}}
\def\ec{\end{center}}
\def\btab{\begin{tabular}}
\def\etab{\end{tabular}}

\begin{document}

\title{\bf LEP~1 - LEP~2: THEORY VERSUS EXPERIMENT\thanks{
Talk given at PADLEP 96, Padova, April 9-11 1996 
and at the One Day Meeting on Open Problems in High Energy Physics,
Frascati, April 10, 1996}}
\author{
Giampiero PASSARINO$^{ab}$,
}
\date{}

\maketitle

\begin{itemize}

\item[$^a$]
             Dipartimento di Fisica Teorica,
             Universit\`a di Torino, Torino, Italy
\item[$^b$]
             INFN, Sezione di Torino, Torino, Italy

\end{itemize}

\noindent
email:
\\
giampiero@to.infn.it

\vspace{4cm}
The status of the standard model is briefly reviewed in the light of the most
recent set of experimental data, with particular emphasis to the limits on
the Higgs boson mass. The search for a light Higgs boson at LEP~2 is also
briefly analyzed.

\newpage

Since new electroweak data from the LEP Collaborations have been made 
available for the winter conferences~\cite{ewwg} we can attempt a partial 
updating of the corresponding theoretical predictions. The first and 
preliminary question that we can address is: what have the theorists been 
doing in 1995-96?

\begin{itemize}

\item Most of the experts have been moving from
MULTI-LOOP TWO-FERMION Physics to TREE-LEVEL FOUR-FERMION Physics

\end{itemize}

Also of some relevance is the following question: what are they planning to 
do next?

\begin{itemize}

\item To move from TREE-LEVEL FOUR-FERMION Physics to either
ONE-LOOP FOUR-FERMION (relevant for LEP~2 and beyond) or to TREE-LEVEL 
SIX-FERMION Physics (relevant for the NLC).

\end{itemize}

Therefore what to say about THEORY versus EXPERIMENT after the recent
upgrading of the experimental accuracy?
The most sensible thing to do is perhaps to present the facts strictly 
separated from the opinions and to be as pragmatic as possible.
In other words we will offer the most advanced standard model (MSM) technology 
and try to find out where we are without any interpretation of the data and 
with no prejudice.

The foundations to discuss estimates of the theoretical error have been set 
by the {\it Working Group on Precision Calculations for the 
$Z$-resonance}~\cite{owg}. 
How do we compare today with the data? This is better illustrated in Table 1
where we have reported for some of the most relevant quantities the 1995 and
1996 experimental error. Moreover we have added the ratio $\Delta_c$ 
between the {\it world average} of the estimate of the theoretical error
and the corresponding experimental one. In the last column of Table 1 we have 
added the most recent evaluation of some of the sub-leading corrections as
computed in ref.~\cite{sl}.

\begin{table}[hbtp]
\begin{center}
\begin{tabular}{|c|c|c|c|c|c|}
\hline
O & '95 err. & '96 err. & '95$\,\Delta_c$ & '96$\,\Delta_c$ & `new' SL \\
  &          &          &                 &                 &          \\
  &          &          &                 &                 &          \\
\hline
  &          &          &                 &                 &          \\
  &          &          &                 &                 &          \\
  &          &          &                 &                 &  $m_t$ at CDF+D0
value        \\
  &          &          &                 &                 &          \\
  &          &          &                 &                 &  $100\,$GeV $\leq$
$M_{_H}$ $\leq$ $300\,$GeV        \\
  &          &          &                 &                 &          \\
  &          &          &                 &                 &          \\
\hline
  &          &          &                 &                 &          \\
  &          &          &                 &                 &          \\
$M_{_W}\,$(MeV)     & 180    & 150     & 2.5$\tmt$ & 3.0$\tmt$ & 
9.5$\tmt$ - 8.3$\tmt$ \\
  &          &          &                 &                 &          \\
$\Gamma_{_Z}\,$(GeV)    & 3.8    & 3.2     & 3.9$\tmt$ & 4.6$\tmt$ &\\
  &          &          &                 &                 &          \\
$R_l$               & 0.04   & 0.032   & 1.0$\tmo$ & 1.3$\tmo$ &\\
  &          &          &                 &                 &          \\
$R_b$               & 0.002  & 0.0016  & 3.3$\tmt$ & 4.1$\tmt$ &\\
  &          &          &                 &                 &          \\
$R_c$               & 0.0098 & 0.0070  & 0.2$\tmt$ & 0.28$\tmt$ &\\
  &          &          &                 &                 &          \\
$A^l_{FB}$          & 0.0016 & 0.0011  & 5.6$\tmt$ & 8.1$\tmt$ &\\
  &          &          &                 &                 &          \\
$A^b_{FB}$          & 0.0038 & 0.00275 & 7.8$\tmt$ & 1.1$\tmo$ &\\
  &          &          &                 &                 &          \\
$A^c_{FB}$          & 0.0091 & 0.0051  & 2.5$\tmt$ & 4.5$\tmt$ &\\
  &          &          &                 &                 &          \\
$sin^2\theta_{eff}$& 0.0004 & 0.0005($A_{LR}$)  & 1.4$\tmo$ & 1.1$\tmo$ & 
1.6$\tmo$ - 1.4$\tmo$\\
  &          &          &                 &                 &          \\
  &          &          &                 &                 &          \\
\hline
\end{tabular}
\end{center}
\caption{Ratios of theoretical uncertainties versus experimental errors.
Data} 
\end{table}

From this table we can easily conclude that the experiments are
closing the gap and the only new piece of calculation is about {\it 
sub-leading} $\Delta\rho$~\cite{sl}. Of course the main question remains

\begin{itemize}

\item Do the experimental data accept the MSM?

\item Are we setting sail for the land beyond the edge of the world, where 
New Physics roam?

\end{itemize}

\noindent
Again we will try to present facts and no opinions, i.e.
to show a {\it poor (standard) man fit}$^*$\footnote{Real fits to be found 
somewhere else} obtained with the help of TOPAZ0~\cite{tz0}. In Table 2 we 
have shown a typical result of a fit with the Higgs boson
mass fixed at $90(400)\,$GeV with the SLD $A_{_{LR}}$ measurement
included(excluded). The uncertainty due to the error on $\alpha(\zm)$ and $m_b$
is fully propagated in the theoretical part of the $\chi^2$. For the given 
$M_{_H}$ the corresponding $\chi^2$ has been obtained by minimizing the 
$\chi^2$ function with respect to $\zm$, $m_t$ and $\alpha_s(\zm)$.
Experimental results will include LEP lineshape + $A_{_{FB}}$ data and their
correlation, LEP asymmetry data, LEP + SLD heavy flavor data and their
correlation, the SLD $A_{_{LR}}$, the $\wm$ measurement and the $m_t$
measurement.

\begin{table}[hbtp]
\begin{center}
\begin{tabular}{|c|c|c|c|}
\hline
O  & Exp. & Theory & Comments\\
\hline
   &      &        & \\
$M_{_H}\,$(GeV) &  --          & $90(400)$ (fixed)   &                    \\
   &      &        & \\
$\chi^2$ & &  $21.5/14(20.2/13)$ & \\
   &      &        & \\
$m_t\,$(GeV)    &  $175 \pm 9$ & $169 \pm 8(176 \pm 8)$ & penalty in the fit \\
   &      &        & \\
$\alpha^{-1}(M_{_Z})$  &  $128.896 \pm 0.09$ & $128.926 \pm 0.099(128.961 
\pm 0.098)$ 
&                \\
   &      &        & \\
$\alpha_s(M_{_Z})$  &  --  & $0.1210 \pm 0.0047(0.1242 \pm 0.0046)$ 
& th. err. not included \\
   &      &        & \\
$m_b\,$(GeV) & $4.7 \pm 0.2$ & $4.67 \pm 0.26(4.67 \pm 0.25)$  &  \\
   &      &        & \\
$\sin^2\theta^e_{eff}$ & $0.23049 \pm 0.00050$ & $0.23137 \pm 0.00027
(0.23188 \pm 0.00029)$ & \\
   &      &        & \\
$\sin^2\theta^b_{eff}$ & $0.2320 \pm 0.0010$ & $0.2326 \pm 0.0002
(0.2332 \pm 0.0002)$ & \\
   &      &        & \\
$R_b$ & $0.2211 \pm 0.0016$ & $0.2159 \pm 0.0003
(0.2156 \pm 0.0003)$ & correlated \\
   &      &        & \\
$R_c$ & $0.1598 \pm 0.0070$ & $0.1723 \pm 0.0001
(0.1724 \pm 0.0001)$ &     "      \\
   &      &        & \\
   &      &        & \\
\hline
\end{tabular}
\end{center}
\caption{Theory versus Experiments(with/without SLD).}
\end{table}

As it is well known the non standard result is $R_b$ and less significantly 
$R_c$), especially in the light of the new data.
For the fun of it let us consider a {\it Transfer Function} $G$, the 
probability of reconstructing the true pair $R_b,R_c$ given the event 
contained the pair of ${\overline R}_b, {\overline R}_c$. For the sake of 
simplicity let us also assume that $G$ is a simple rotation ($\theta$) in the 
$R_b,R_c$ plane. We assume to believe in a light Higgs($100\,$GeV) and 
perform a fit to $\zm$ $m_t, \alpha_s$ and $\theta$ and obtain

\begin{eqnarray}
m_t &=& 171 \nll
\alpha_s(\zm)  &=& 0.1211 \nll
\theta &=& 1.27^o \nll
\end{eqnarray}

\noindent
giving

\begin{eqnarray}
R_b &\mid&  0.2159 \to 0.2195 (1.0-\sigma),  \nll
R_c &\mid&  0.1723 \to 0.1675 (1.1-\sigma),
\end{eqnarray}

\noindent
which is not bad since they are $1.0\,\sigma$ and $1.1\,\sigma$ away from
the experimental value. New Physics should be compatible with rotating
$b-c$ of about $1^o$.

\noindent
Back to {\it Orthodoxy} it is perhaps opportune to re-iterate that so far
no {\it realistic} calculation exists for the $\barb b$ cross section
which should take into account as much as possible the experimental
setup chosen to extract $R_b$. 
Every realistic calculation should interface the two-fermion final state with 
the four-fermion one since by now we know how to include the {\it exact} 
$\barb b \barc c$ cross section into $\sigma(\barb b)$ and $\sigma(\barc c)$. 
This is most likely not going to account for deviations but we need the 
real thing and we should not go for anything less. Thus we have shown in
Fig. 1 the ratio $\sigma(\barb b \barc c)/\sigma(\barb b)$ as a function of
$\sqrt s$ as computed by TOPAZ0-WTO~\cite{wto}.

The next question will be of course about supersymmetry, i.e. how do we 
compare the MSM with the MSSM? 
Data are still changing a little and a new plot of DATA/MSSM is not 
yet available. However things do not change dramatically in the data from 1995
to 1996 and the old conclusions remain valid to a very large extent. Thus 
$50\%$ of the discrepancy on $R_b$ will go away with MSSM but nothing will be 
gained in $R_c$ where however the data have been changing a little moving
towards the theoretical predictions.

All of this is really peanuts, the  edge of the world is the {\it Higssland}
since a light Higgs is most likely supersymmetric while a heavy Higgs means 
troubles for almost everybody. We have attempted several fits but before
discussing the outcome we would like to stress the following facts:

\begin{itemize}

\item The main problem is to understand when the $\chi^2$ shape as a function
of $M_{_H}$ is unstable with respect to {\it normal} fluctuations of the
experimental data in the large $M_{_H}$ tail.

\item Very stringent bounds on $M_{_H}$, i.e. much less than $500\,$GeV, look 
more like a symptom of the clash between SLD and LEP.

\item The $\chi^2$ shape depends {\it heavily} on the introduction of
penalty functions which constrain $m_t$ and/or $\alpha_s$. 

\item $\chi^2_{min}(M_H)$ has an unnatural tendency to be in the 
{\it forbidden} region, thus requiring the unnatural introduction of yet 
another penalty function.

\end{itemize}

\noindent
In all fits we have used the most recent electroweak data~\cite{ewwg} with or 
without the exclusion of some sub-set. We have fixed the Higgs boson mass, fully
propagated the uncertainty on $alpha(\zm)$ and $m_b$ in the fit and
determined $\zm$, $m_t$ and $\alpha_s(\zm)$. After that the $\chi^2(M_{_H})$
curve or the $\Delta\chi^2(M_{_H})$ curve has been constructed and the
$95\%$ C.L. one-sided upper bound has been determined ($\Delta\chi^2 = 2.7$).

In Fig. 2 we have shown the effect of including the theoretical error
in the fit, at least according to TOPAZ0. Therefore the two solid lines 
account for different options in treating higher order corrections as 
equivalent treatments of resummation techniques, momentum transfer scale for 
vertex corrections and factorization schemes. As depicted here the one-sided 
upper bound on $M_{_H}$ has become quite less than the {\it mythical} 
$1-1.5\,$TeV of some year ago. Fig.3 is telling us what the effect will be of 
lowering the experimental error on the top quark mass or on $\wm$. Clearly is 
not only a question of reducing the experimental error but also the effect of 
the central value is of some relevance for the goodness of the fit, which is 
not of the highest quality anyway. In Fig. 4 we have attempted to understand 
the effects of single measurements on the determination of the Higgs boson 
mass. It emerges that the exclusion of $R_b,R_c$ but expecially of $A_{_{LR}}$ 
will move up $M_{_H}$ considerably. The only statement that should be made 
today is that while we are well below the $1\,$TeV wall (with this set of 
data) it is still premature to give something more precise than a vague 
estimate $M_{_H} \leq 400-500\,$GeV at $95\%$ of confidence level. In 
particular we find somehow unsatisfactory the large jump in the upper bound 
when a {\it single} measurement is removed from the set of data.

Even more important we have analyzed the question of stability of the
$\chi^2$ high $M_{_H}$ tail. The procedure is standard, by going at the minimum
of the $\chi^2$ we use that value of $M_{_H}$ to compute the electroweak
observables according to the MSM (as seen by TOPAZ0). This set of numbers,
hereafter termed {\it fixed theory} (FT), is subsequently used instead
of the experimental data (E) for a new fit. Thus the difference between the
solid and the dashed line in Fig. 5 gives terms linear in T-FT (E-FT drops
out of the $\Delta\chi^2$) which are {\it noise}, i.e. normal fluctuations.
As a matter of fact the goodness itself of the fit is not completely 
satisfactory. Whenever a penalty on $m_t$ is included, accounting for the
CDF+D0+UA2 constraint, we see that the two minima of $\Delta\chi^2$ for
T-FT or T-E are slightly shifted. This is due to the fact that while repeating
the fit for T-FT the theory prefers to adjust first the constraint of the
penalty paying a much lesser price on the rest of the residuals. This again
we interpret as suggesting some caution in establishing a very precise upper
bound on $M_{_H}$ in the region below $400-500\,$GeV. Thus our only conclusion
will be $M_{_H} \leq 500\,$GeV from the present set of data (average among
the four LEP Collaborations) and the present correlation matrix.

If the Higgs boson is in the range of LEP~2 we should stop worrying about 
{\it Tails\&Fits} and start to understand how it will look like in a {\it real}
environment. Already a large amount of work has been done in this 
direction~\cite{nyr} and here we will briefly summarize the present 
theoretical situation by using the result of WTO. All the results have been 
obtained by including the $H \to gg$ channel in the total Higgs width. 
This treatment will therefore evolve $\alpha_s$ to the
scale $\mu = \hm$, evaluate the running $b,c$-quark masses and compute

\begin{eqnarray}
\Gamma_{_H} &=& {{G_G\hm}\over {4\,\pi}}\,\left\{ 3\,\left[ 
m_b^2(\hm) + m_c^2(\hm)\right]\,\left[ 1 + 5.67\,{\alpha_s\over \pi}
+ 42.74\,\lb{\alpha_s\over \pi}\rb^2\right] + m_{\tau}^2\right\} +
\Gamma_{gg},  \nll
\Gamma_{gg} &=& {{\gf\hm^3}\over {36\,\pi}}\,{{\alpha_s^2}\over {\pi^2}}\,
\lb 1 + 17.91667\,{\alpha_s\over\pi}\rb.
\end{eqnarray}

\noindent
Finally the Higgs boson {\it signal} is multiplied by

\begin{eqnarray}
\delta_{_{QCD}} &=& 1 + 5.67\,{{\alpha_s}\over {\pi}} + 
42.74\,\lb {{\alpha_s}\over {\pi}}\rb^2,  \nll
\alpha_s &=& \alpha_s(\hm).
\end{eqnarray}

\noindent
In Fig. 6 we have shown the $e^+e^- \to \mu^+\mu^-\barb b$ total cross section 
as a function of $\sqrt{s}$ for three possibilities: no Higgs, $M_{_H} = 80\,$
GeV and $M_{_H} = 100\,$GeV. At LEP~2 a large fraction of events will be
of the type $\nu\barnu b\barb$ and we have shown the corresponding cross
section (summed over all neutrinos) in Fig. 7 with a simple set of kinematical 
cuts, $|M_{\mu\mu}(M_{\nu\nu}) - \zm| \leq 25\,$GeV and $M_{\barb b} \geq 50\,$
GeV.
To show the potentialities of the dedicated electroweak codes we have
presented in Fig. 8 and 9 the $M_{b\barb}$ distribution for $e^+e^- \to
\mu^+\mu^-\barb b$ at two energies, $175\,$GeV and $190\,$GeV and two
values of $M_{_H} = 80,100\,$GeV. Actually in Fig. 8 and 9 we have shown the
ratio

\begin{equation}
R = {{M_{_H}}\over {\sigma}}\,{{d\sigma}\over {dM_{b\barb}}}.
\end{equation}

\noindent
Finally in Fig. 10 we have compared the invariant mass distribution
$d\sigma/dM(b\barb)$ with the corresponding histogram reporting the number
of events/$0.5\,$GeV.

\vskip 35pt
\noindent
Special thanks are due to Manel Martinez, Dorothee Schaile, Robert Clare,
Wolfgang Hollik and Paolo Gambino.


\newpage

\begin{figure}[htbp]
\begin{center} \mbox{
\epsfysize=19cm
\epsffile{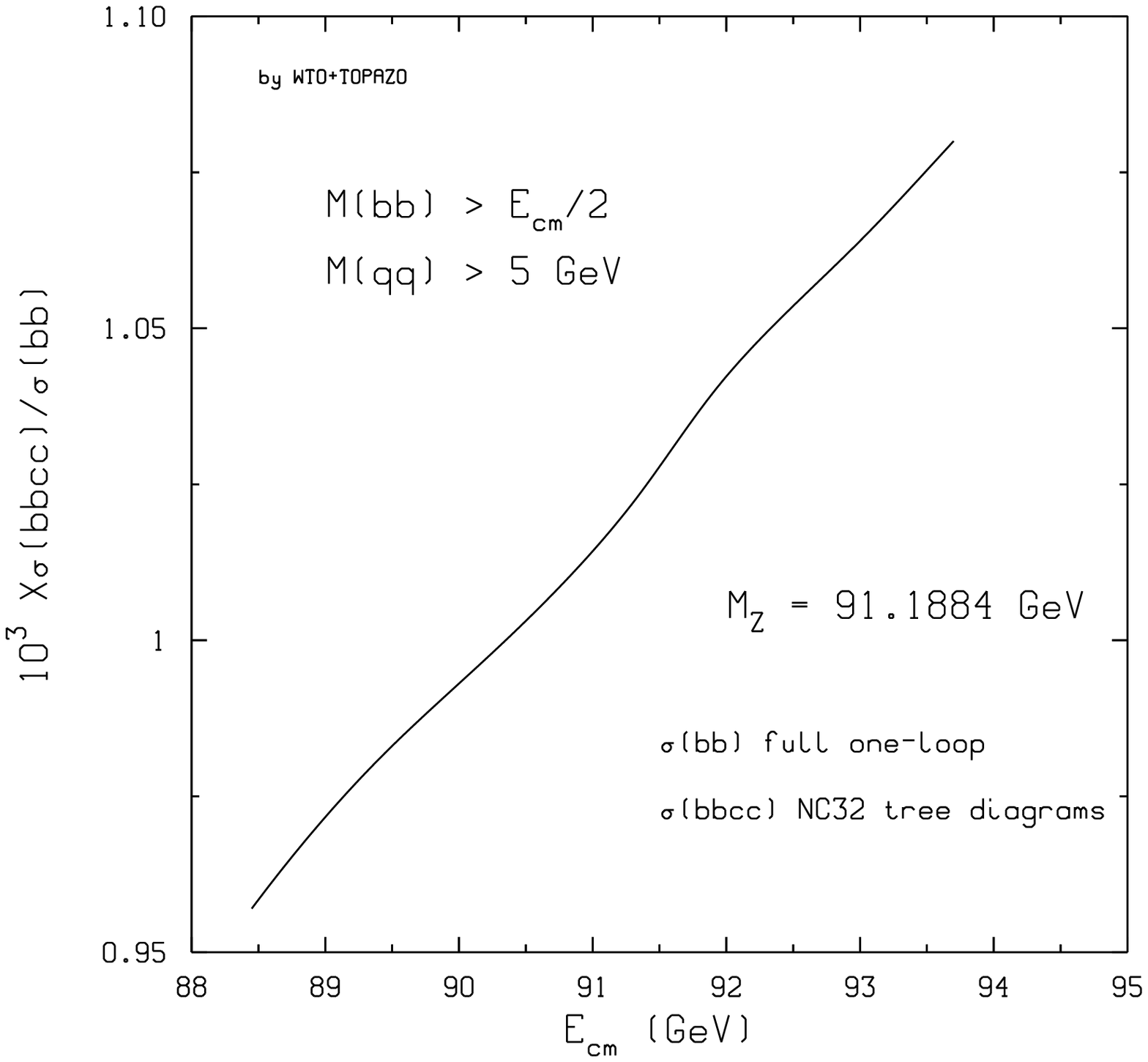}}
\end{center}
\caption[
The ratio $\sigma(\barb b \barc c)/\sigma(\barb b)$ at LEP~1 energies.
]{
{
The ratio $\sigma(\barb b \barc c)/\sigma(\barb b)$ at LEP~1 energies.
}  }
\end{figure}

\newpage

\begin{figure}[htbp]
\begin{center} \mbox{
\epsfysize=19cm
\epsffile{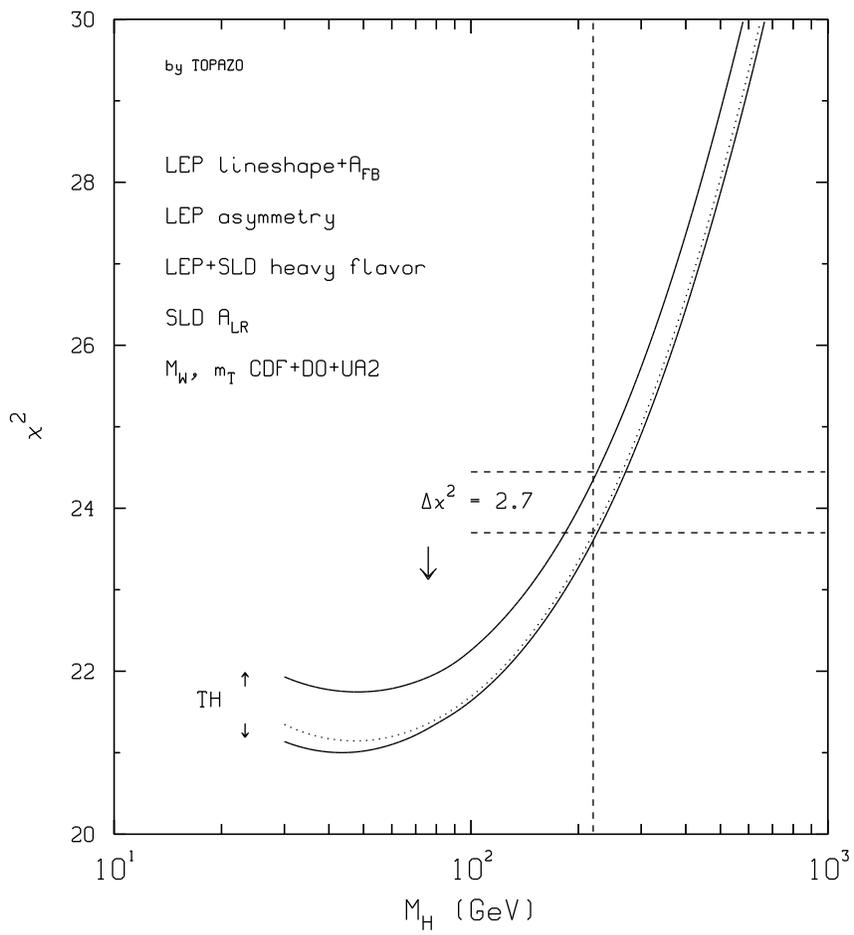}}
\end{center}
\caption[
The $\chi^2(M_{_H})$ curve inclusive of the estimate of the theoretical error.
]{\label{fig2}
{
The $\chi^2(M_{_H})$ curve inclusive of the estimate of the theoretical error.
}  }
\end{figure}

\newpage

\begin{figure}[htbp]
\begin{center} \mbox{
\epsfysize=19cm
\epsffile{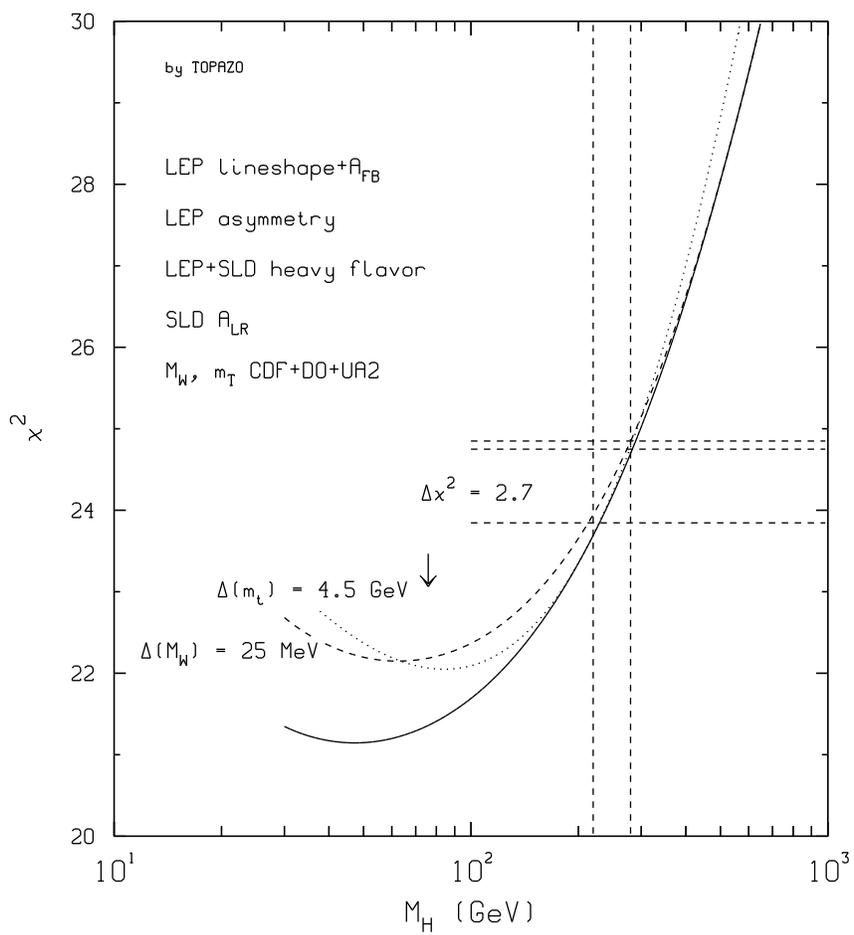}}
\end{center}
\caption[
The $\chi^2(M_{_H})$ curve with projected experimental errors on $m_t$ and 
$\wm$.
]{\label{fig3}
{
The $\chi^2(M_{_H})$ curve with projected experimental errors on $m_t$ and 
$\wm$.
}  }
\end{figure}

\begin{figure}[htbp]
\begin{center} \mbox{
\epsfysize=19cm
\epsffile{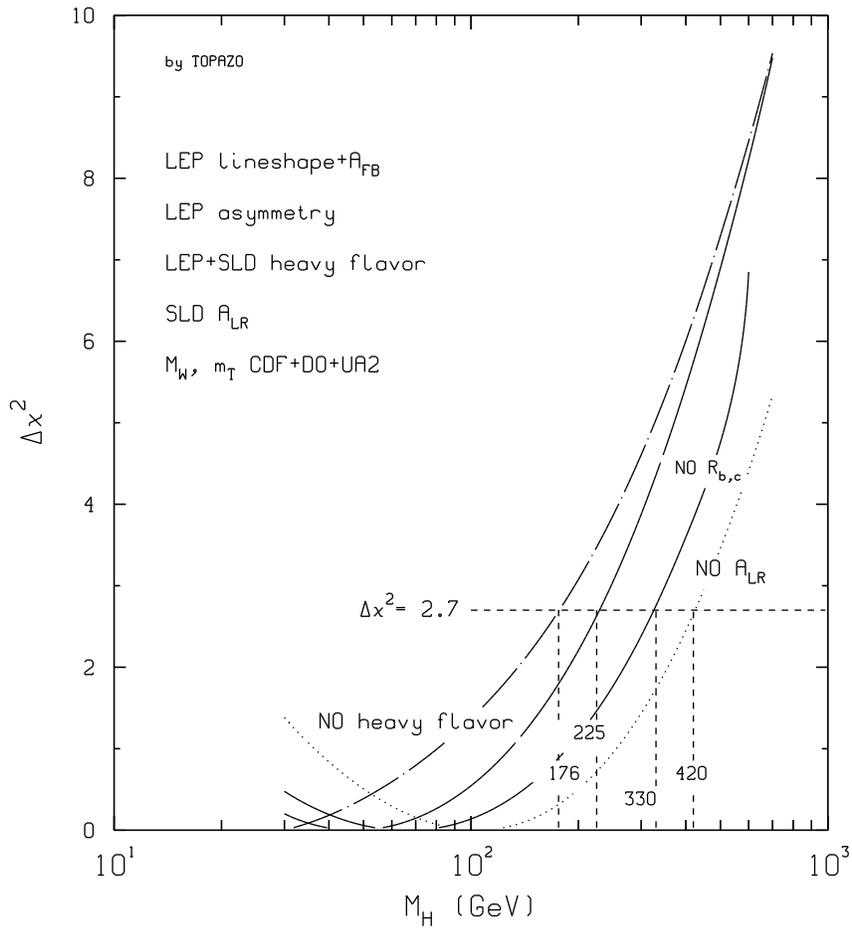}}
\end{center}
\caption[
The $\chi^2(M_{_H})$ curve with the exclusion of some sub-set o experimental 
data.
]{\label{fig4}
{
The $\chi^2(M_{_H})$ curve with the exclusion of some sub-set o experimental 
data.
} } 
\end{figure}

\begin{figure}[htbp]
\begin{center} \mbox{
\epsfysize=19cm
\epsffile{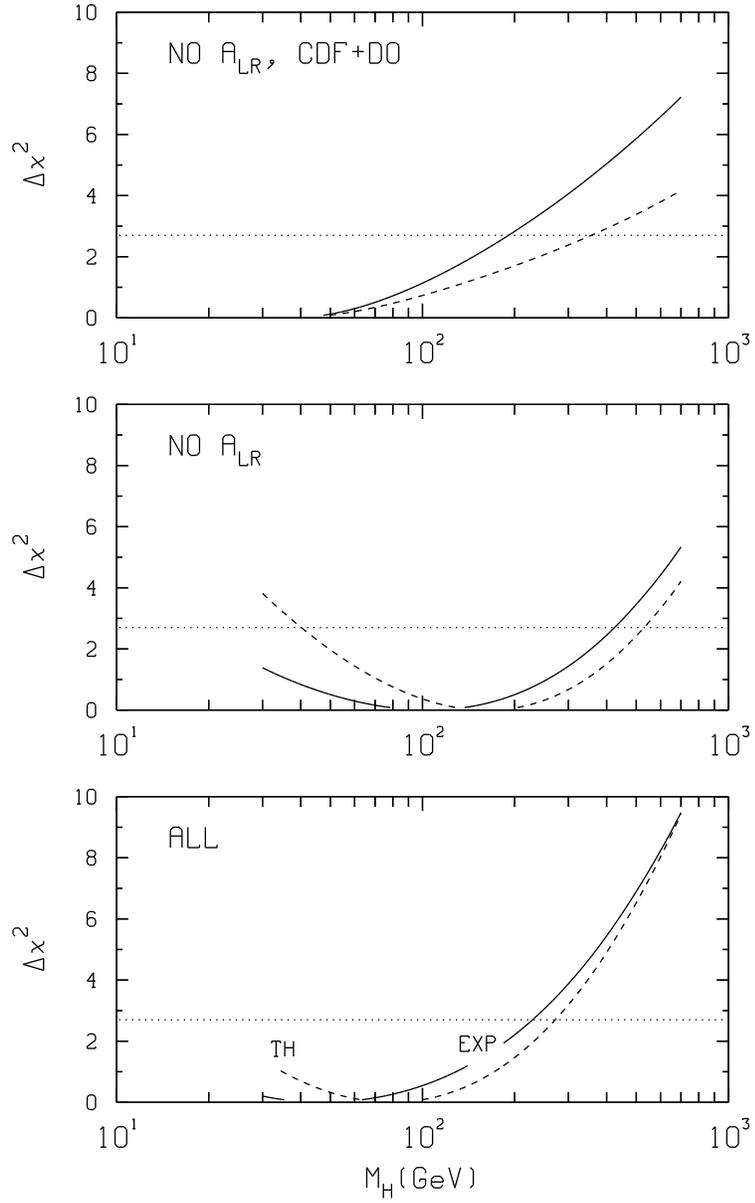}}
\end{center}
\caption[
The normal fluctuations in the $\Delta\chi^2(M_{_H})$ curve. 
]{\label{fig5}
{
The normal fluctuations in the $\Delta\chi^2(M_{_H})$ curve. 
}  }
\end{figure}

\newpage

\begin{figure}[htbp]
\begin{center} \mbox{
\epsfysize=19cm
\epsffile{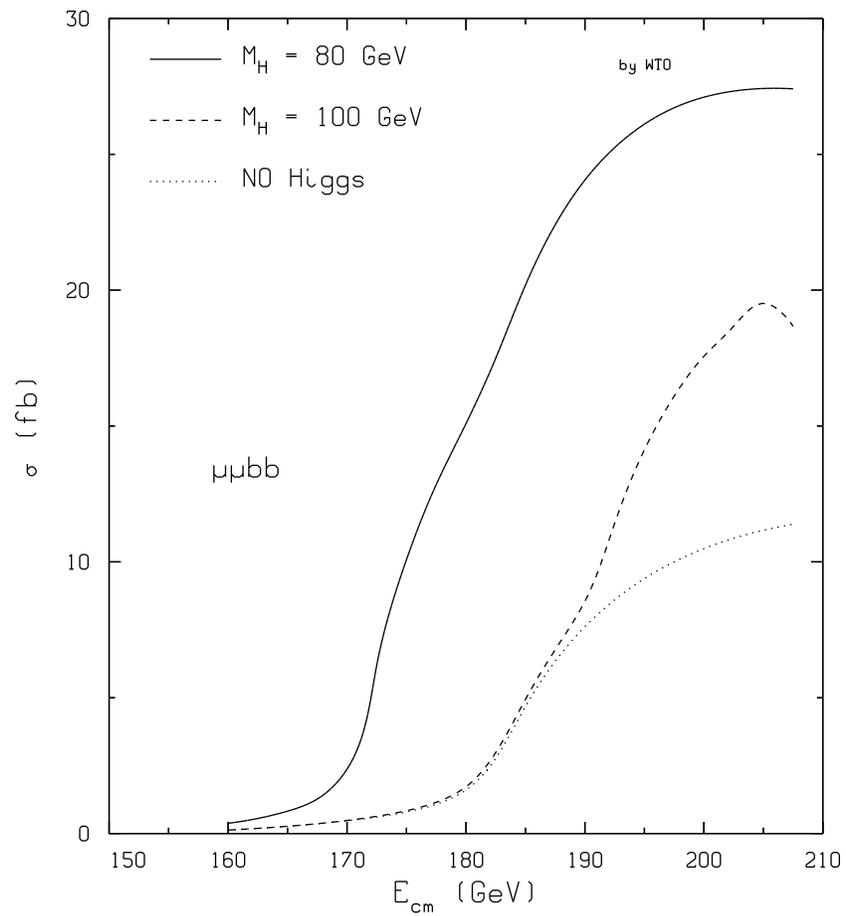}}
\end{center}
\caption[
The cross section for $e^+e^- \to \mu^+\mu^-\barb b$.
]{\label{fig6}
{
The cross section for $e^+e^- \to \mu^+\mu^-\barb b$.
}  }
\end{figure}

\newpage

\begin{figure}[htbp]
\begin{center} \mbox{
\epsfysize=19cm
\epsffile{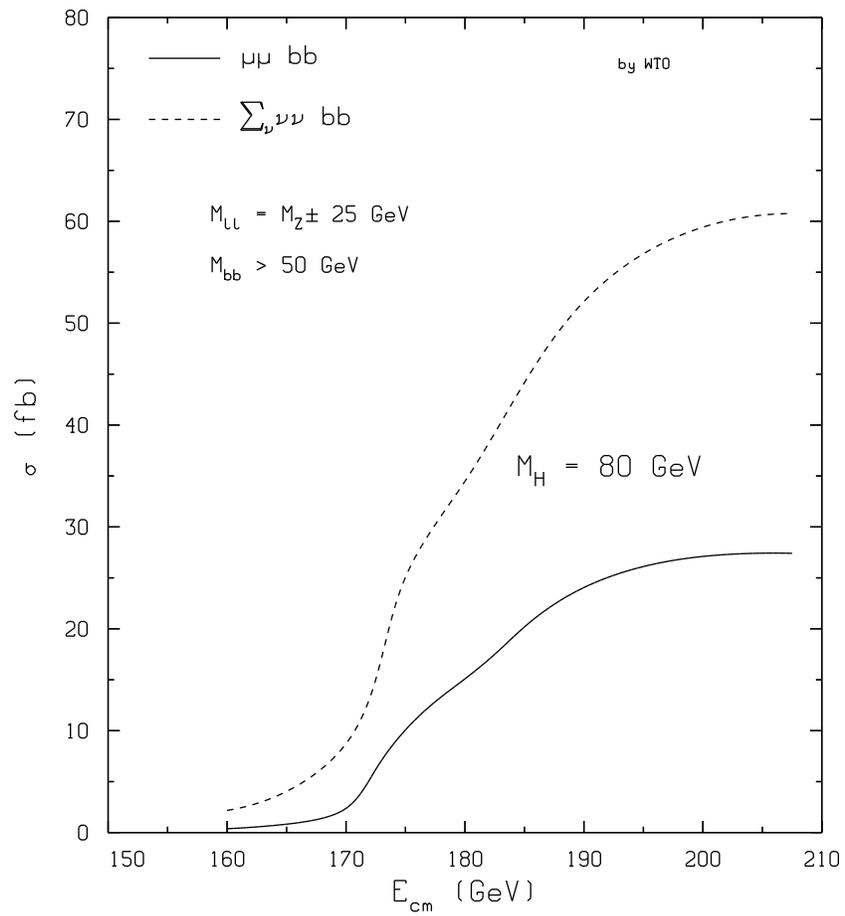}}
\end{center}
\caption[
The cross section for $e^+e^- \to \mu^+\mu^-(\sum\barnu\nu)\barb b$.
]{\label{fig7}
{
The cross section for $e^+e^- \to \mu^+\mu^-(\sum\barnu\nu)\barb b$.
}  }
\end{figure}

\newpage

\begin{figure}[htbp]
\begin{center} \mbox{
\epsfysize=19cm
\epsffile{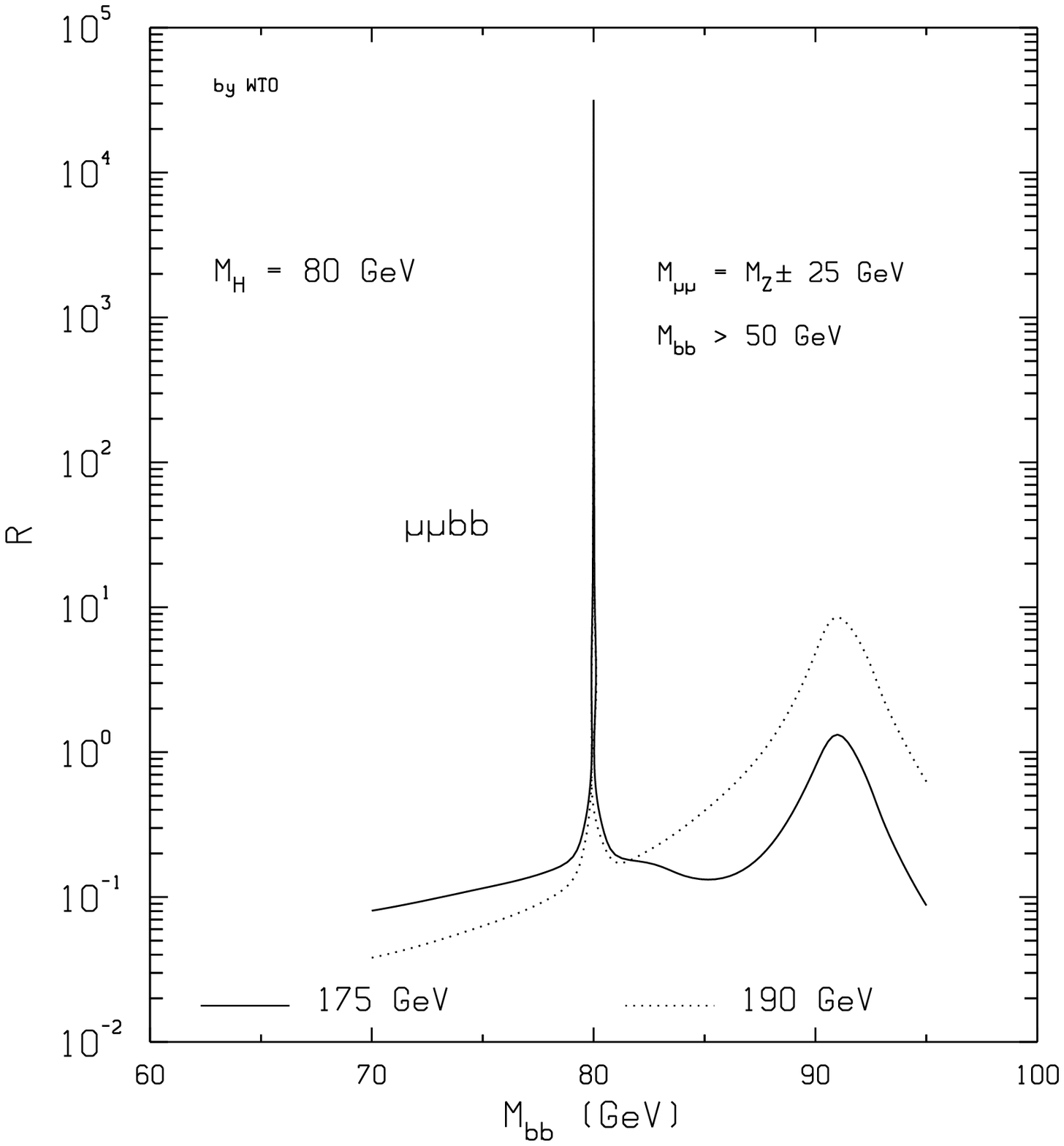}}
\end{center}
\caption[
The $M_{\barb b}$ distribution for $e^+e^- \to \mu^+\mu^-\barb b$ at $M_{_H} 
= 80\,$GeV.
]{\label{fig8}
{
The $M_{\barb b}$ distribution for $e^+e^- \to \mu^+\mu^-\barb b$ at $M_{_H} 
= 80\,$GeV.
}  }
\end{figure}

\newpage

\begin{figure}[htbp]
\begin{center} \mbox{
\epsfysize=19cm
\epsffile{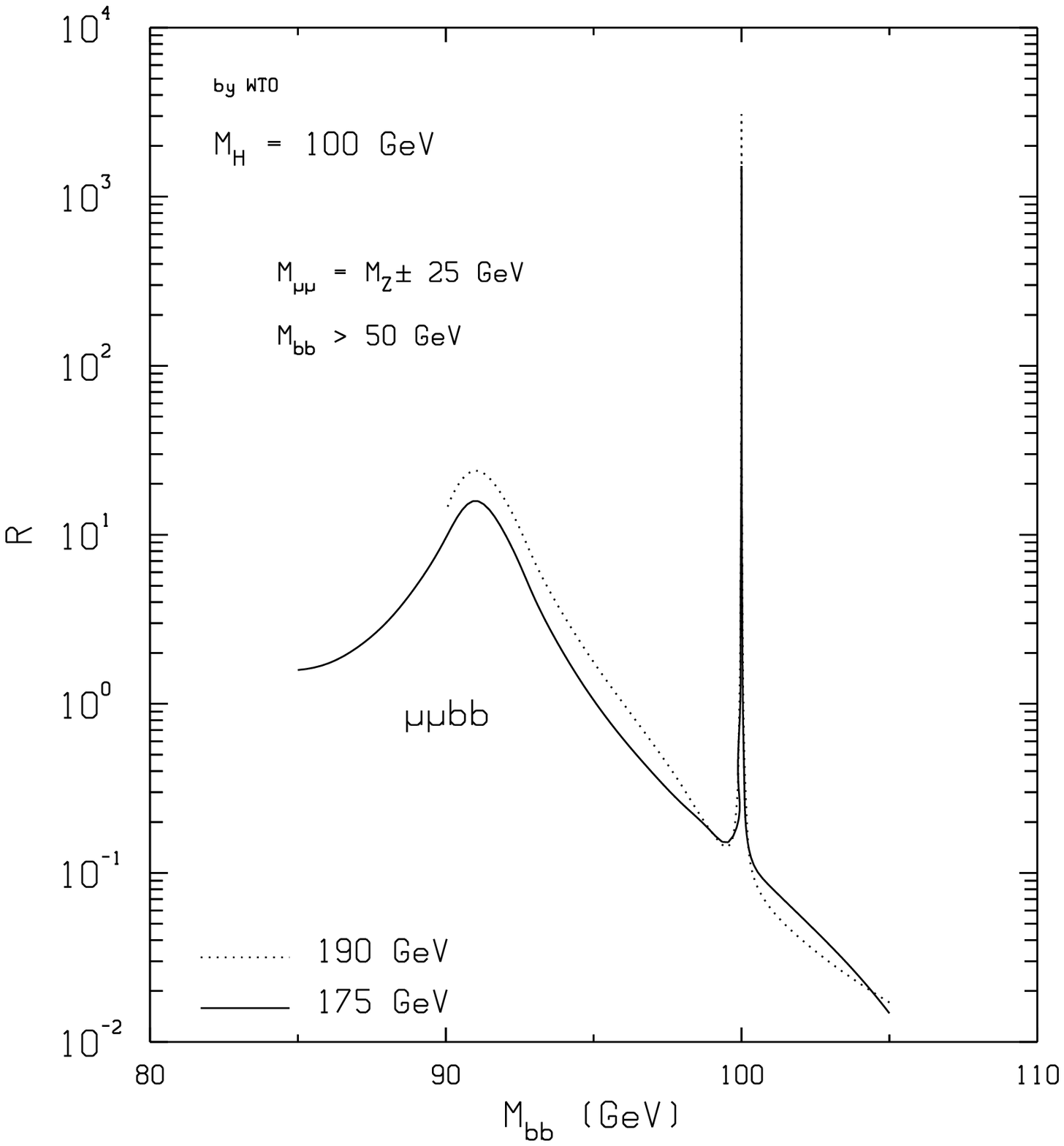}}
\end{center}
\caption[
The $M_{\barb b}$ distribution for $e^+e^- \to \mu^+\mu^-\barb b$ at $M_{_H} 
= 100\,$GeV.
]{\label{fig9}
{
The $M_{\barb b}$ distribution for $e^+e^- \to \mu^+\mu^-\barb b$ at $M_{_H} 
= 100\,$GeV.
}  }
\end{figure}

\newpage

\begin{figure}[htbp]
\begin{center} \mbox{
\epsfysize=19cm
\epsffile{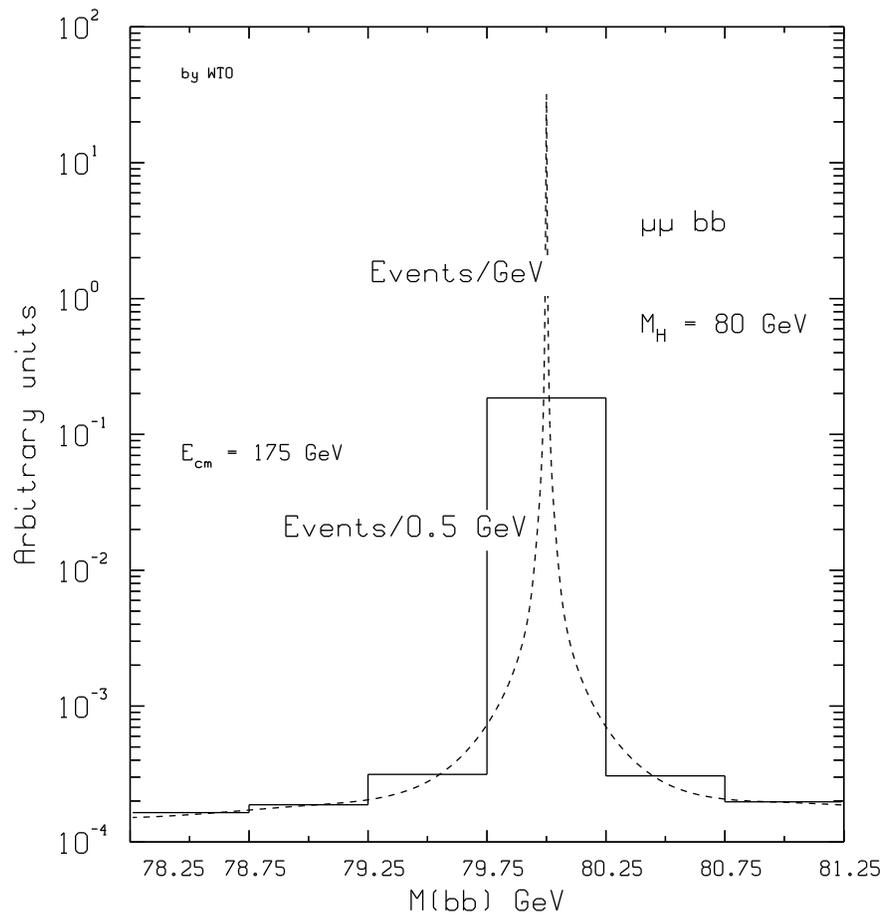}}
\end{center}
\caption[
The $M_{\barb b}$ distribution for $e^+e^- \to \mu^+\mu^-\barb b$ versus 
events/$0.5\,$GeV. 
]{\label{fig10}
{
The $M_{\barb b}$ distribution for $e^+e^- \to \mu^+\mu^-\barb b$ versus 
events/$0.5\,$GeV. 
}  }
\end{figure}

\end{document}